# INTELLIGENT USER INTERFACE IN FUZZY ENVIRONMENT

Ben Khayut[1] , Lina Fabri[2] and Maya Abukhana[3]

[1]Department of R&D, IDTS at Intelligence Decisions Technologies Systems, Ashdod, Israel
ben_hi@hotmail.com
[2]Department of R&D, IDTS at Intelligence Decisions Technologies Systems, Ashdod, Israel
lina.fabri@gmail.com
[3]Department of R&D, IDTS at Intelligence Decisions Technologies Systems, Ashdod, Israel
maya.kh111@gmail.com

## ABSTRACT

*Human-Computer Interaction with the traditional User Interface is done using a specified in advance script dialog "menu", mainly based on human intellect and unproductive use of navigation. This approach doesn't lead to making qualitative decision in control systems, where the situations and processes cannot be structured in advance. Any dynamic changes in the controlled business process (as example, in organizational unit of the information fuzzy control system) make it necessary to modify the script dialogue in User Interface. This circumstance leads to a redesign of the components of the User Interface and of the entire control system. In the Intelligent User Interface, where the dialog situations are unknown in advance, fuzzy structured and artificial intelligence is crucial, the redesign described above is impossible. To solve this and other problems, we propose the data, information and knowledge based technology of Smart/ Intelligent User Interface (IUI) design, which interacts with users and systems in natural and other languages, utilizing the principles of Situational Control and Fuzzy Logic theories, Artificial Intelligence, Linguistics, Knowledge Base technologies and others. The proposed technology of IUI design is defined by multi-agents of a) Situational Control and of data, information and knowledge, b) modelling of Fuzzy Logic Inference, c) Generalization, Representation and Explanation of knowledge, c) Planning and Decision-making, d) Dialog Control, e) Reasoning and Systems Thinking, f) Fuzzy Control of organizational unit in real-time, fuzzy conditions, heterogeneous domains, and g) multi-lingual communication under uncertainty and in Fuzzy Environment.*

## 1. INTRODUCTION

### 1.1. Analysis

*Data*, *information* and *knowledge* are one of the main *components* in the world development. *Their* correct use leads to the adoption of *relevant* decisions and *effective dialog control* through the *Intelligent Multi-agent User Interface* (IMAUI) in *Fuzzy Control System* (FCS).
The traditional User Interface (UI) is using a 'Menu" principle to interact with their users (analysts, experts, managers). Each of them spends an *unproductive* time to *navigate* the menu dialog instead of solving their day to day tasks. They can be more *productive* in interacting with the UI in more natural and less predefined way, where the UI *itself* should find and process the required data, information or knowledge to support their tasks. Changes in business process of organizational unit entail changes in the script menu dialog. This leads to a permanent and continuous *redesign* of the UI.
In these circumstances, since the technology doesn't support the business process need, it is impossible to use the UI in *real-time*. To avoid such a *discrepancy* it becomes necessary to use the *principle* of Situational Control [6] in the design of IUI. Given the *uncertainty* of the environment, as well as increased *complexity* of business process and technologies, the use of Fuzzy Logic [1] and Artificial Intelligence becomes *necessary*. A multi-lingual Processor (LP) should be used to support Human-Computer Interaction (HCI) to incorporate a business process into control system. In order to ensure *functionality* of the IUI in *fuzzy environment* in the paper it is suggested to use of Fuzzy Logic Inference [13], Generalization and Explanation of knowledge [15], Dialog Control [11] and other methods. Obviously, the considered *smart/ intelligent IUI* should be able to support a variety of *subject areas*, *natural* and other *languages.*
The solution of these tasks in this article is focused on the *systematic* and *linguistic* approach of *modelling*, *planning* and *controlling* of *linguistic* and *subject area* data, information, knowledge, *fuzzy logic inference, dialog control* and others, by mapping the *objectives* and *constraints* in *fuzzy environment*.
The *novelty* of the technology which designs *IMAUI* consists of:

- *Modelling* and *situational fuzzy control* of data, information and knowledge for implementing an *automatic* fuzzy inference and finding a correct, accurate, timely and adequate *decision,* taking into account a current *situation* and impact of *fuzzy environment*.
- Using of resulting *decision*, criteria and purpose for providing of *modelling, planning* and *control* of the business process in the fuzzy environment.
- Converting and deriving images, concepts, meanings from *natural languages* in various subject areas and *serializing* them into the *bases* of data, information and knowledge.
- Use of these *bases* for *multi-lingual* human - machine *communication* using methods of dialog control, generalization and explanation of knowledge in the *intelligent fuzzy control system.*
- Use of *properties* of a) *atomicity* of data, b) *relational relationships* of information, c) *figurativeness* and *associative connectedness* of knowledge for their *integration* and *aggregation.*
- Using *methods* of *wisdom, intuition, behaviour* and others to obtain decision of high quality and precision.

The analysis of the *state* of scientific research in the field of design *IUI* showed that the directions and methods of implementation are related mainly to the *functionality* of the *agents*, controlled by *them*. In this context, we will hold a brief of comparative analysis of the *functionality* of *agents of* the *IUI*, offered by us and other authors.
In [12] is given an interesting overview of the approach to the problem of *Fuzzy Control*. Our approach *differs* from the mentioned, the fact that in addition to proposed methods of formalization we taking into account the principles of *situational control*, *artificial intelligence* and others. This allows realizing fuzzy control in situation, which *unknown in advance*. At the same time, we have developed *modelling* techniques [17] based on the *managed* data, information and knowledge [10] that allows finding relevant solution with the *desired accuracy* in the *circumstances*. This accuracy is implemented using *intelligent agents* of analysis, decision-making, planning and others [22] by using the values of fuzzy membership function.
In [18] is represented linguistic approach for solving decision problems under linguistic information using Multi-criteria decision making, linguistic modeling, aggregation and linguistic choice functions methods on base of rank ordering among of the alternatives for choosing the best of them.
The main difference between our systematic approach and the proposal is a:

- *Generalized notion of linguistic variable* of Fuzzy Logic, by which we evaluate and take into account not only the *morphological*, *syntactic* and *semantic*, but also, *behavioural*, *psychological* and other *aspects* of the terms (atomic units) of Natural Language (NL).

- *Situational Fuzzy Control in Fuzzy environment*, by which we control not only information, but also data, knowledge, decisions, agents and others.
- *Decision-making* process is based not only on using the *rank* for estimation of the alternatives, but also on *automatic* Fuzzy Logic *Inference, Planning, Control* of alternatives, situations and other units.
- *Multi-lingual interaction*, *generalization*, *explanation*, *serialization*, *storage* and *actualization* of *knowledge* in fuzzy conditions, *heterogeneous subject areas*, where the situations are *unknown in advance*, fuzzy structured and not clearly regulated.

In [20] are considered adjustable autonomous agents that possess partial knowledge about the environment. In a complex environment and unpredictable situations these agents are asked the *help of human* on base of the model, called HHP-MDP (Human Help Provider MDP) and requests, which *are set* in advance.

The comparative analysis of these and other works, associated with our work, showed, that there is no *integrated, systemic and linguistic* approach to the problem of *situational fuzzy control* in a *fuzzy environment*, including the techniques of situational control of fuzzy data, information and knowledge, modeling, planning, decision-making, dialog control and situational fuzzy control of the *organizational unit*, based on the achievements of *Fuzzy Logic*, *Situational Control, Artificial Intelligence*, *Linguistics, Knowledge base technologies* and others.

In this article, we present the results of our studies and the approach to the design of *IMAUI* using our developed methods and tools.

## 1.1. Terminology

*Data* is organized in the memory and are perceived by the person or machine as facts, numbers, words, symbols, lines and other items of information. They are not related to each other and are found in texts, pictures and other maps of reality.

*Information* is a group of *related data*, organized in the memory that respond to the questions of "who", "what", "where", "when" and others.

*Knowledge* is the *image* or domain *model*, extracted from *information* and organized in memory, which in *itself* are interpreted, structured, linked, associated, transformed, compared, upgraded, activated, analyzed, deduced, built, serialized and so on, in real time. The mentioned *image* or *domain* should respond to questions "why" and "how", consider the impact of environment and specificity of subject area, satisfying the criteria and purpose of existence.

*Wisdom* is a method of perceiving reality and achieving a unique solution (answer) on the basis of intelligence, archival knowledge (experience), principles and *inference* in a *certain situation.*

*Intuition* is a method of perceiving reality and a achieving a unique solution (answer) on the basis of intelligence, archival knowledge (experience) principles and *unique inference* in an *extreme situation.*

*Modelling decisions* is defined as construction of a new conceptual situation and a state of controlled units (fuzzy data, information, knowledge, inference and others), which meets the criteria and purposes of the information system in fuzzy environment. The purposes are functions of the information system.

*Planning decisions* is defined as a use of modelling results to create a sequence of alternative decisions that will match to the situation and the state of information system in the subsequent stages of management of the organizational unit.

*Decision-making* is defined as a process of modelling *fuzzy logic inference* [13] for selection the relevant decision from limited number of alternative decisions, obtained during the *planning decisions*.

*Reasoning* is a method (process) of *making inferences* from *body* (premises) of data, information and knowledge.

*Fuzzy Control* is the process of using the modelling results of planning and decision-making in fuzzy environment, in order to implement a control action on the units (data, information, knowledge, decisions, organizational unit and others) to shift them and their control system to a new state that matches a specified criterion.

Under the *fuzzy logic inference* [13] we mean *procedure* for determining the vector of internal and external *output* fuzzy variables $b_k^i \in V_k^m$ using a *new* vector of the values of the *input* fuzzy variables $a_k^i \in U_k^m$, which *transforms* the *IMAUI* in its *new state*. This *procedure* is implemented on the extensional, intentional and reformative levels of modelling knowledge [14].

Under the *Dialog Control* we means the process of presenting partners (IUI-Human or IUI-IUI) of common commands (questions) to each other and providing by them targeted actions (issuing replies) relevant to the subject of the dialogue and the situations in which it occurs. The process allows using the *natural* and\ or other *languages*).

*Intelligent fuzzy control system* is understood as a *knowledge based* system, which is reliably electronic *autonomous* system, and which a) operates at a high-level operating system b) connected to the Internet, d) executes a native or cloud-based applications, e) analyzes the collected data, information and knowledge, and e) realizes the human-machine functions for solving problems in fuzzy environment.

*Traditional UI* is an interface, which is working on base of "menu" scenarios and is not autonomous.

The *IUI* is understood by us as knowledge *based* system, which controls the *intelligent functionality* of all agents in fuzzy control system considering the situation and conditions of uncertainty and the impact of them surrounding Fuzzy Environment.

According Wikipedia the *Organizational Unit* $O^U$ (Figure 2, Figure 4) represents a single organization with multiple units (departments) within that organization.

The *business process* is an activity or set of activities in *organizational unit* $O^U$ that will realizes a specific organizational goal.

*Subject area* understood by us as branch of knowledge and technologies, where the organizational units are functioning.

*Environment* is the surrounding reality, consisting of organizational units, information systems, robots, agents, agent systems and so on, which interact with each other under the influence of the environment.

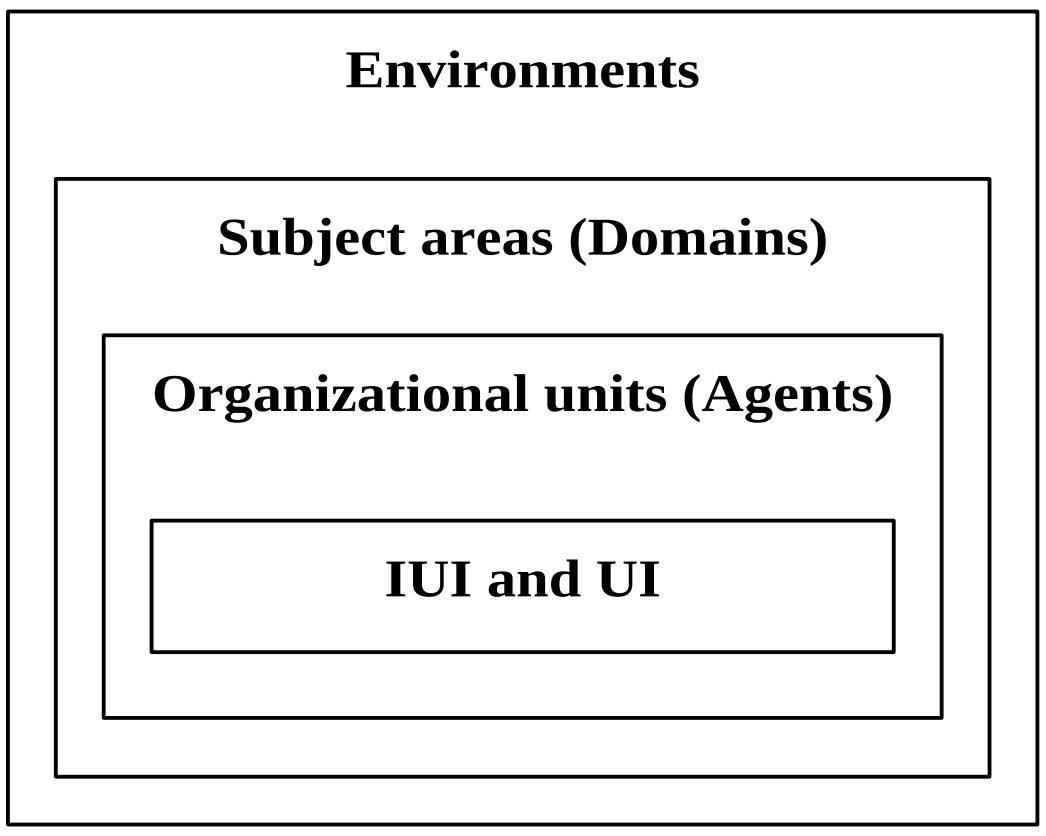

Figure 1. IUI and UI environments

Figure 1 depicts nesting of the above-defined concepts.

*A multi-agent system (MAS)* [19] is a computational system where *agents* cooperate or compete with others to achieve some individual or collective task.

*Agent* is a *real-world* or *artificial* entity, which is a *person* (in the first case) and an *object* (in the second case), and which are capable of performing some *action* or *service* or otherwise, *interacting* with other entities.
Thus, the above defined and implemented in the computer concepts are *agents*.
Combining, nesting and integrating the agents into the groups according to their *objectives* and *functional* features turn them into an *Intelligent Multi-agent System* in the paper.
*Fuzzy Control* provides a formal methodology for representing, manipulating, and implementing a human's heuristic knowledge about how to control a system [12].
According Wikipedia the *Fuzzy Control System* is a control system based on fuzzy logic – a mathematical system that analyzes analog input values in terms of logical variables that take on conditions values between 0 and 1, in contrast to classical or digital logic, which operates on discrete values of either 1 or 0 (true or false, respectively).

# 2. CONTENT

## 2.1. Conception

The main purpose of the IUI function is to *facilitate* the user to perform analysis, planning, decision-making, management and coordination of organizational unit. Under the *facilitating* in IUI is understood the *full use* of *intelligence* of the system to perform these and other functions by interacting with it in *natural* and/ or other *language*. In this regard, the *function* of the IUI should be based on the perception, processing and synthesizing knowledge in real time.
Given the fact that these processes occur in unexpected situations and *Fuzzy Environment* (people, robots, nature, space, hardware, software, other information and agent systems, and so on) becomes necessary to use the theories *Situational Control* [6], *Fuzzy Logic* [1] and other methods.

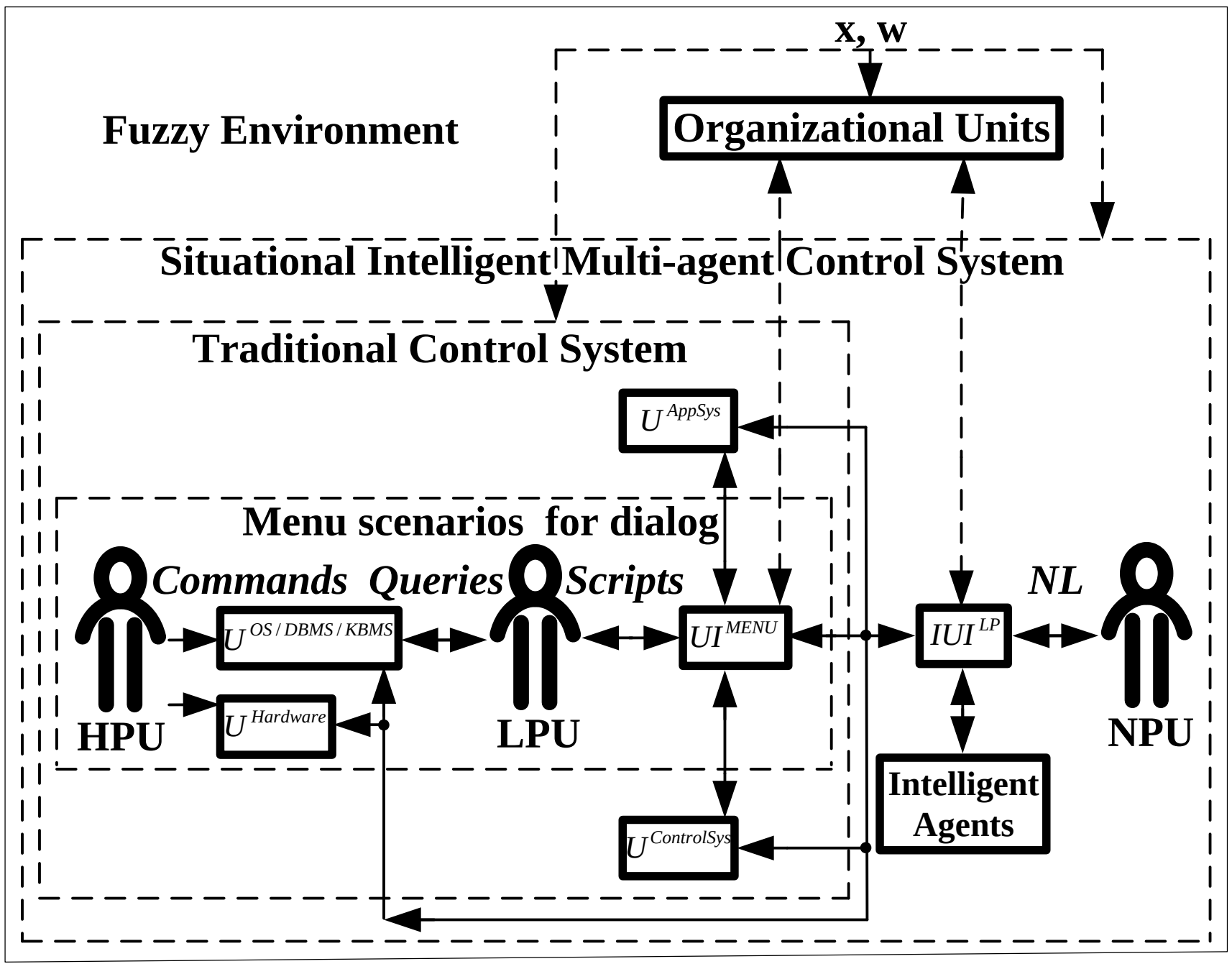

Figure 2. Human Interaction using Smart/ Intelligent and Traditional UI.

The paper proposes the *creation technology* of the *IMAUI* that *integrates* the achievements of Situational Control, Fuzzy Logic, Linguistics, Artificial Intelligence and others for *realization* of the following *functionalities:*

- *To use of models* of representation of *linguistic* and *subject area* data, information and knowledge in Fuzzy Environment.

- *Situational fuzzy* Modelling, Decision-making, Control and Planning in conditions of the absence, incompleteness, vagueness and ambiguity of knowledge.
- *Fuzzy Logic Inference.*
- *Generalization* and *Explanation* of knowledge.
- *Fuzzy Dialog Control.*
- *Reasoning and Systems Thinking.*
- *Linguistic Processor* for *Multi-lingual* interaction in *Fuzzy Environment*.

The main features of the proposed technology of design *IMAUI* lies in the use of methods:

- Situational control of data, information and knowledge and implementation by it of *automatic* processes of inference, making the right decisions, generalization and explanation knowledge, dialog control, planning and management of organizational unit.
- Creation *LP*, which is a part of the *IUI* and enabling people to interact with him in Natural Language *without* "menu" dialog.

The *group of* highly *professional users* (HPU) (Figure 2) includes users, which were trained in the use and maintenance of hardware, software, networks, data, information and knowledge bases.

A *group of less professional users* (LPU) in (the mentioned above resources) are developers: system analysts, application programmers, testers, operators and others.

The *group of not professional users* (NPU) is composed of *experts* in their field, who use the *functionality* of information system to solving their *functional* tasks. This group includes the *decision makers*, *analysts, experts, consultants, managers* and other experts in their subject area.

$UI^{MENU}$ is the *user interface* of dialogue menu in Traditional Information System, with whom interacts user LPU using *script dialogue language*. This user also interacts with interfaces $UI^{OS/DBMS/KBMS}$, using the *query* language.

$UI^{OS/DBMS/KBMS}$ is a *group of UI*, used by users HPU for system support the Operation Systems (*OS*), Data Base Management Systems (*DBMS*) and Knowledge Base Management Systems (*KBMS*) using *commands languages.*

$U^{AppSys}$, $U^{ControlSys}$ are respectively, *Application* and *Control* systems, which realize the *functionalities* of the considered Traditional Control System and other systems in it.

$IUI^{LP}$ is *IUI* with built in LP and with whom interacts the user NPU using *natural language*. This *interface* is connected with interfaces $UI^{OS/DBMS/KBMS}$, $UI^{MENU}$ and with systems $U^{AppSys}$, $U^{ControlSys}$ for the use of the existing functionalities of other information systems. Thus, *IMAUI integrates* the *functionality* of existing *traditional UI* and *control systems* to solve *simple* problems and its *intellectual* capabilities to solve *complex* problems in *fuzzy environment.*

$U^{Hardware}$ are *hardware resources*, supported by user HPU.

The interface $IUI^{LP}$ interacts with the organizational entities, performing all functionality of *FCS* by returning to user NPU the results in the required form. Similar work is done by the user LPU, which interacts with the *organizational units* through the interface $IUI^{LP}$.

*Intellectualization* of labor of the LPU and HPU users *is not* considered in this paper and will be the subject of further study. These studies involve the introduction of artificial intelligence in hardware, software and networks, and the inclusion of these groups in the NPU group.

## 2.2. Methods

### 2.2.1. The models and methods of representation of linguistic and subject data, information and knowledge in Fuzzy Environment.

For *storing* and *representation* of *subject* and *linguistic* data, information and knowledge we *integrated*, *generalized* and used the language features of *RX-codes*, *syntagmatic chains*, *semantic networks, and universal semantic code* and *frames* languages [21]. These languages are used to establish *conformity* between *relational* level of knowledge representation and *logic-pithy* level of representation of data, as well as their in a formalized *general* presentation of knowledge using procedures of *pithy* information processing.
We described the *apparatus* for constructing these models and establish this *conformity* on *declarative*, *procedural* and *transformative* levels of representation data, information and knowledge in the form of aggregated and integrated *semantic network of computing frames* [10].
In [10], [13] were extended the concept of a *linguistic variable*, formal and *semiotic models* with using the *principle* and method of *situational control* [6] by taking into account of the accepted methods of representation, organization, integration, processing and synthesis of data, information and knowledge [3-5], [7,8], [14]. Through the use of *fuzzy sets* theory and *situational control* model were defined *linguistic* and *thematic* units, attributes, corteges and *dictionary entries* in linguistic and thematic relational bases of data, information and knowledge.
These *models* define a conceptual means of presenting and structuring of data, information, knowledge in fuzzy environment, and are also used for modeling, planning, decision-making and fuzzy control [22] in *IUI*.

Thus, the *intellectuality* of the *Data, Information and Knowledge Control System* consists in providing of interaction of decision-maker with consultants and experts (last among themselves) in order to organize *dialogue* between them in a *natural language using IUI*.

### 2.2.3. Situational Fuzzy Modelling, Decision-making, Control and Planning in conditions of the absence, incompleteness, vagueness and ambiguity of knowledge.

The *feature* of solving the problem of *modelling* knowledge is related to the construction of multilingual linguistic processor, modelling of fuzzy logic inference, generalization and explanation of knowledge, and reasoning using the methods of situational data, information and knowledge control in the particular domain of organizational control. For this purpose: a) developed *fuzzy model* of modelling, representation and synthesis of knowledge, b) generalized the notion of *linguistic variable* of fuzzy logic, c) formalized and constructed *semiotic model* of situational control of thematic and linguistic knowledge, and d) developed a method of modelling knowledge, based on built *semantic network of computational frames*.

The *principle* of the method of *modelling knowledge* [14] is *consists* in processing of aggregate *semantic network* of *concepts* using *knowledge units*, where each *generalized linguistic variable* is modelled in a particular *segment of accumulation* of hypothetical *polyhedron* through the *dictionary entry* and *cortege* in the *thematic* and *linguistic* knowledge bases.

The *vertices* of said *polyhedron* are considered as a result of integration, aggregation and meaningful interpretation of the various *points of view* on a particular *aspect* of the *real world*, expressed on natural language. In this case, the *knowledge modelling* processes are implemented on *natural language* and *machine* levels by switching from one to another (i.e. from *verbal* to *numeric* ratings) is being realized on base of the model of situational knowledge control using of knowledge modules and LP.

In order to control an organizational unit it is required to now its structure, the purpose of its existence and its control criteria [6].

The task becomes more complicated when there is a need to control organizational units in real time, in situations unexpected in advance, using variety of natural languages and subject areas.
In these circumstances, arises a problem of decision making in fuzzy environment [2] based on the data, information and knowledge.

The solution to this problem is implemented by a) methods of modelling, planning and controlling of linguistic and subject area data, information, knowledge [10], [22], fuzzy inference [13], decision making, reasoning and others, b) mapping the objectives and constraints in fuzzy environment [17]. The methods of Fuzzy Modelling, Planning, and Decision-making were described in [22].

### 2.2.4. Fuzzy Logic Inference

Given the complex character of functioning of the *IUI*, its design is impossible without the use of theories situational control [6], fuzzy sets [1] and the proposed above of models of representation, synthesis, modelling, planning and management of data, information and knowledge.

Therefore, the modeling method of fuzzy inference can be applied to data control system, satisfying the following principles [13], [10]:

- All information about the data, information and knowledge (about the organizational unit) may be communicated to the control system as a set of phrases of Natural Language.
- Control model is fundamentally should be open and never ends the creation of the final formal model.
- Description of the data management (information, knowledge) process is possible in the form of natural phrases and \ or another language.

In these circumstances, the proposed modeling method of fuzzy inference is implemented by a system of situational data control and displays a *linguistic approach* to the problem. The method allows realizing the inductive and deductive inference in natural language in integrated subject areas, based on incoming fuzzy fragments (parcels) of the language.

To do this, we used the heuristic *algorithms*, *methods* of wisdom, intuition, behaviour and other algorithms and methods that invoke the modules of modelling of data, information and knowledge. The algorithms and methods uses generalized linguistic variables, fuzzy sets, rules and facts (situations), previous decisions and their subsets (segments), extracted from bases of data, information and knowledge to obtain relevant decision.

The description of the *fuzzy logic inference* is given in [13] and is understood as a *process* for determining the vector of internal and external *output* fuzzy variables $b_k^i \in V_k^m$ using a *new* vector of the values of the *input* fuzzy variables $a_k^i \in U_k^m$, which *transforms* the system in its *new state* by means of *fuzzy matching* in a *fuzzy modeling relation*, through *compositional mapping rules*, objectives, constraints, associations and *measures of opportunities*. This *process* is implemented by means of modeling knowledge [14] and is realized as a *procedure* of *commutative* mapping of *verbal* and *numerical* values of the logic functions $\mu^q_{B_{Lijk}}$ ( $R^{X_{i,j,k}}_{i,j,k}$ ) and the relationship $R^{X_{i,j,k}}_{i,j,k}$ in a *fuzzy relationship of modelling*, presented in [6], [13] by the diagram (Figure 3).

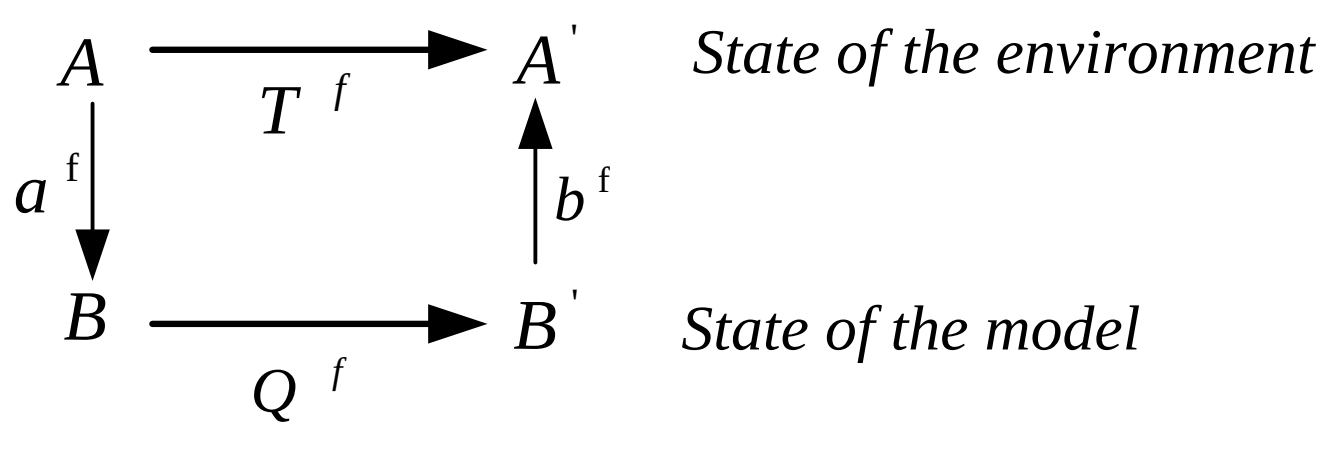

Figure 3. Diagram of a fuzzy relationship of modelling.

The $a^f$, $b^f$ are respectively maps, defined by matrices of verbal and numerical estimates of functions $\mu^q_{B_{Lijk}}$ ( $R^{X_{i,j,k}}_{i,j,k}$ ) and relationship $R^{X_{i,j,k}}_{i,j,k}$ .

$A$, $A'$ - are states of environment.

$B$, $B'$ - are models associated with these states in the environment.

$T^f$, $Q^f$ - are respectively transformations in environment and in model of situational control.

$f$ - are frames $\Phi^q_{L_{ijk}}$ of modeling.

Under the *fuzzy matching* we mean the *frame* $\Phi^q_{L_{ijk}} : U^m_k \leftrightarrow V^l_k$ that displays the matching of base fuzzy sets $U^m_k$ and $V^l_k$ in the *procedure* of inference, through the *compositional mapping rules*

$$\mu^q_{B_{Lijk}}(X) = \mu^q_{A_{Lijk}}(X) \circ \Phi^q_{L_{ijk}} \qquad (1),$$

where:

$\mu^q_{B_{Lijk}}(X)$, $\mu^q_{A_{Lijk}}(X)$ are respectively, the resultant and initial *membership functions* in the considered *generalized fuzzy relation* $R^q_{L_{ijk}}$ ;

$X = \{x_{ijk}\}$ is a vector of the discrete domain of definition;

*o* is sign of the compositional mapping;

$\Phi^q_{L_{ijk}}$ - is *fuzzy matching* in the procedure of inference;

$A_{Lijk}$, $B_{Lijk}$ are heterogeneous multi-dimensional fuzzy sets;

$L_{ijk}$ - are heterogeneous *distributive lattices* of measurement intervals of the domains of considered *membership functions*;

*q* defines levels of representation of the data, information and knowledge, respectively on *RX – codes*, *Universal Semantic Code* and *Semantic Frames* [21] *levels*.

*i, k, m = (1, n*).

Then

$$\mu^q_{B_{Lijk}}(X) = \bigvee_{p \in P} \left( \bigwedge_{k \in K} \mu^q_{A_{Lijk}}(X) \right) \bigwedge_{j \in I} \overset{q}{Poss}(a_{ijm} / a'_m) \qquad (2)$$

defines the mentioned above *compositional mapping rules* of the fuzzy inference by using *measures of opportunities* $\overset{q}{Poss}(a_{ijm}/a'_m)$ on $q$ levels of modeling data, information and knowledge.

It should be note that $\overset{q}{Poss}(a_{ijm}/a'_m)$ defines the *measure of the possibility*, that the *composite concept* in the *external* representation is characterized by its values $a_{ijm}$ of a fuzzy set $A_{Lijk}$ in the *internal* representation.

### 2.2.5. Generalization and Explanation of knowledge

The task of *generalization* of knowledge is reduced to finding the *target* (*unique*) situation $Q_l$ of data, information and knowledge by using of their current situation $Q_j$ and process of *control* them on base of *model* [15].
The *decision*, correspond to the found situation of the data, information and knowledge, shifts them from the current situation $Q_j$ into a new $Q_l$.
This *decision (action)* determines the *impact rules I* on data (information and knowledge), which must be met in the overall situation $S_i$, so that they and control system would correspond to the new (changed) situation $Q_l$.
*The target function* in the model of generalization and explanation of knowledge defines the purpose of control of data, information and knowledge. The purpose may be the stirring up of processes of modelling, decision-making, planning, control, generalization and explanation knowledge and other.

The process of generalization is seen as a search for the objective function (situation) by modeling the knowledge and fuzzy inference using the semantic structure of data, information and knowledge.

The concepts are interpreted as a group of lexical units of a language (of generalized linguistic variables, economic indicators) designed to refer to the facts, phenomena and of elements using of dissimilar verbal and numerical estimates ($\alpha$ - cuts) on the internal and external levels of representations by humans and computers.

The formalized model of generalization and explanation of knowledge (data, information) in the IUI in [22] is represented by ratio

$$< A^N, K^R, \mu_R^T(\mu_R^\phi), (S_i : Q_j \overset{x,u,w}{\Rightarrow} Q_l : I) > \qquad (3),$$

where:

$A^N$ is a model of relevant decision, received from a large number of alternatives *A*, which contains *linguistic variables* (generalized situations);

$K^R$ is a formal system of data, information and knowledge modelling (generalized *knowledge module*), which is working on base of *rules* and *relations* in the *segments* of *semantic network of computational frames*;

$\mu_R^\phi$ and $\mu_R^T$ are, respectively, the *values (estimates)* of fuzzy logic functions in fuzzy *knowledge control,* and *generalization and explanation* processes;

According of [6], [13] the $S_i : Q_j \overset{x,u,w}{\Rightarrow} Q_l : I$ determines the *elementary act* of control in the processes of modelling and selecting the relevant decision, that transforms the control system in the new situation $S_i$, which characterizes its new state $Q_l$ after the state $Q_j$ was shifted to $Q_l$.

The process of *explanation* knowledge is realized as *inversed* process of *generalization* knowledge.

### 2.2.6. Dialog Control

In this paper, we propose an approach to the control of the dialogue [11] in IUI using the modeling and control data, information and knowledge [10], fuzzy inference [13], generalization and explanation knowledge [15], *Reasoning* and *Systems thinking* and others.

Together with these control systems, Dialogue is controlled by *planning* system, which uses the model, created by interpreter of dialogue $U^D$ (Figure 4) and sub systems of fuzzy inference $U^I$, generation and explanation of knowledge $U^G$, reasoning and systems thinking $U^Z$. Those sub systems (*agents*) provide processing and forming the input and output messages of natural language in knowledge management system using *bases* of *subject area* and *linguistic* data, information and knowledge.

The $IUI^L$ is an *IUI,* witch contains the LP. Together with the Manager the $IUI^L$, $U^D$, $U^I$, $U^D$, $U^Z$ are representing the Decision-making system. The $U^A$ (analyst, reviewer) and $U^E$ (expert, approver) are intelligent subsystems, which are support the decision-making process.

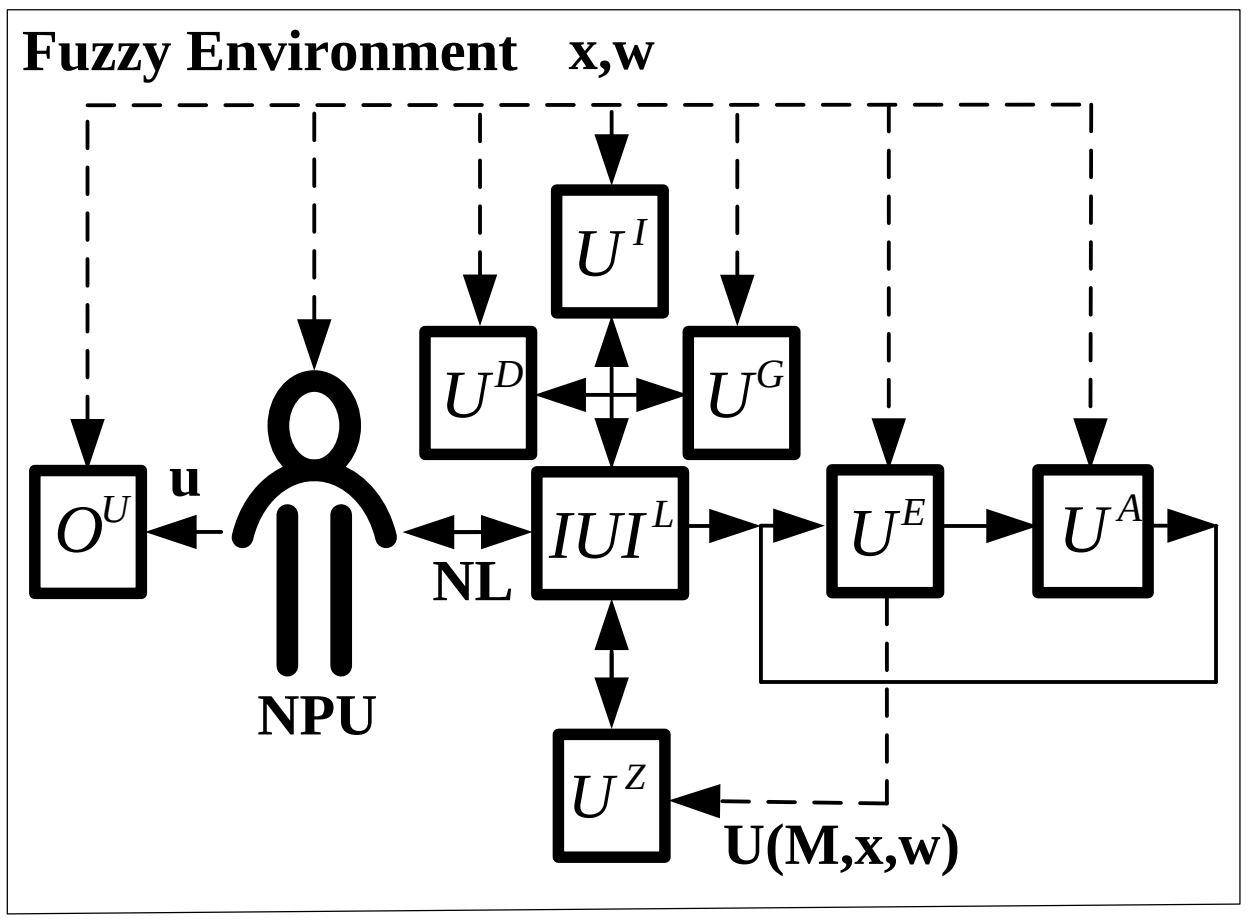

Figure 4. Dialog Control in human interaction.

The task of dialog control is formalized in [22] by the ratio

$$< A^N, K^R, \mu_R^D(\mu_R^T(\mu_R^\phi)), (S_i : Q_j \overset{x,u,w}{\Rightarrow} Q_l : I) > \qquad (4),$$

where $\mu_R^D$=*f* ($\mu_R^T(\mu_R^\phi)$, u, x, w) - is a fuzzy logic function, which depends on the other mentioned logical functions $\mu_R^T$, $\mu_R^\phi$ and managed of environmental influences *u, x, w* on the organizational unit with the target to make relevant decisions on control, respectively, at the *intentional*, *extensional* and *transformative* levels of their modeling.

### 2.2.7. Reasoning and Systems Thinking

The *essence* of the methods of *Reasoning* and *Systems Thinking* consists in *extracting* of relevant data, information and knowledge from texts (speech) in different *natural languages*, within dissimilar *subject areas* to implement situational *fuzzy control* of data, information and knowledge with a view of representing them in *knowledge bases* and organizing on basis of them *processes* of reasoning and thinking *under uncertainty*. On the basis of *thinking* are formed, *associated* concepts, which are *related* them with the appropriate symbols of language in a particular subject area. Taking into account the *scheme of reasoning* and its *modelling* suggested by [16] we adhere to the *concept*, where the *main objective* of knowledge representation is to promote the *modelling* process of *Fuzzy Logic Inference* using methods of *fuzzy control* and *actualization* of data, information, knowledge in *knowledge bases*.

Thus, the *Reasoning* is considered in this paper as a *process* of gaining *new knowledge* and *parcels* on basis of the mentioned a) *logic inference process*, b) generalization and explanation knowledge, and c) dialog control, which were situational fuzzy *controlled* by the processes of fuzzy *modelling* and *control* of data, information and knowledge, according the *objective* under uncertainty in Fuzzy Environment.

The *formalized task* of *Reasoning and Systems Thinking* can be presented as

$$< A^N, K^R, \mu_R^Z(\mu_R^D(\mu_R^T(\mu_R^\phi))), (S_i : Q_j \overset{x,u,w}{\Rightarrow} Q_l : I) > \quad (5).$$

In this case,

$\mu_R^Z(x) = \vee(\mu_R^D(x) \wedge \mu_R^T(x) \wedge \mu_R^\phi(x) \wedge Poss(a/a^\alpha)$ is a *composite rule* of reasoning and thinking in *IMAUI*, where the rule $\mu_R^Z(x)$ is interpreted as a desired fuzzy *logic function*, which identifies the *disjunction* of *conjunctions* of modulated *assessments* of *values* of characteristic fuzzy logic functions $\mu_R^D(x)$, $\mu_R^T(x)$, $\mu_R^\phi(x)$ and *conformity measure* $Poss(a/a^\alpha)$.

**2.2.8. Linguistic Processor for Multi-lingual interaction in Fuzzy Environment**

The users interact with *IUI* using natural language, which is processed by *LP*. The *LP* contains the *Interpreter* and *Synthesizer* agents.

Their implementation involves the *functionality* [3, 5] of:

- Providing meaningful *machine translation* for identification concepts.
- The *adequacy* of the mapping meanings of the concepts expressed by the term in a particular language in the context of a particular subject area.
- Opportunity to endow a specific term by attributes of *grammar*, *logic*, *semantics*, *pragmatics*, *psychology* and others in accordance with its meaning in a particular subject area and the context of use.
- Implementation of the "*understanding*" by means the algorithms of synthesis output expressions of natural language based on *logical-semantic* characteristics of intra-linguistic representations of meaning.
- Resolution *disambiguation* expressions used languages by sampling lexical and semantic characteristics and vocabulary of languages and conduct on the basis of their lexical and semantic analysis to determine the semantic matching of input and output language equivalents.
- The possibility of implementing a system of automatic machine dictionaries and of thesauri in linguistic databases with conceptual connection to the subject databases and knowledge.

Implementation of Machine Translation terminological phrases includes the blocks of:

- Logical-semantic, grammatical analysis of input combinations and the identification of concepts of its output equivalents.
- Prior authorization of lexical ambiguity on input phrases using conceptual codes.
- Extraction of grammatical information.
- Grammatical ambiguity resolution based on the stems of input combinations.
- Extracting logical information.
- Final resolution of lexical ambiguity.
- Resolution of lexical and grammatical ambiguity of input and output combinations of stems.
- Logical and semantic disambiguation of input and output terminological expressions.

- Identification of concepts and the formation of natural language expressions of output terminological phrases.

The *LP* realizes transformation target's actions, which are expressed in NL and other languages. The method of realization of the *LP* is represented in [3].

## 3. Conclusions

The proposed methods and technology are oriented for design and development of *autonomous Smart/ Intelligent IUI*, which will be operated in *fuzzy environment*, interacting with people and other systems and agents in *different languages* and dissimilar *subject areas* , where the situations and factors of influence on the *IUI*, control system and organizational unit cannot be determined and structured in advance.
A distinctive feature of this approach is to use an *IUI* to help users to implement modelling and control of fuzzy linguistic and subject data, information, knowledge, alternative solutions, objectives and constraints in order to find accurate, relevant and right decisions, which are suitable to the situation in *Intelligent Fuzzy Control* systems.
The results of this work focused on the creation *multi-agent IUI* for *multilingual interaction* in *autonomous, situational, intelligent, multi-agent information control* systems of robots, unmanned production and spacecrafts and others, functioning in the fuzzy environment and unforeseen situations in advance.

**Authors**

*Ben Khayut* graduated MSc in *Mathematics* at the University of Chisinau and PhD in *Mathematical Cybernetics* (completed studies) at the Institute of Mathematics of the Academy of Sciences of the Moldova in the field of *applied mathematics*, *computer science*, *artificial intelligence*, *fuzzy logic*, where he defended the pre-thesis "*Data Control in decision-making on natural language*". Has experience (over 30 years) in *design* and *development* of *Information Control systems* in industry using *linguistic* and *subject area* data, information and knowledge bases, and *algorithms* of machine translation as researcher, designer, programmer, project manager, using modern scientific methods, software and hardware. Have 25 scientific papers. In 2007 he founded the *Intelligence Decisions Technologies Systems* (IDTS).

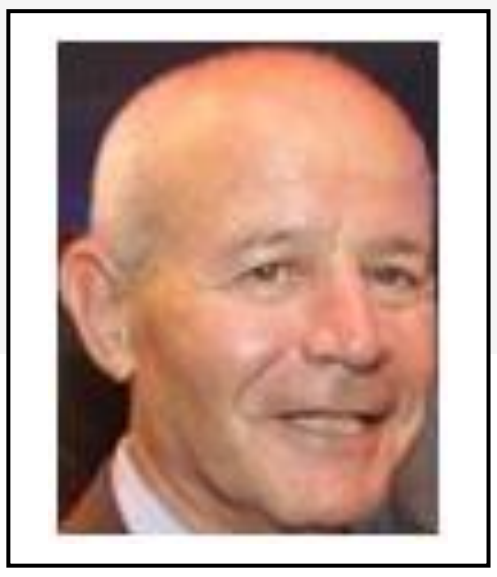

*Lina Fabri* graduated *Bachelor*, *Economics and Management* at the College of Business Management. Have 14 years of successfully leading and executing *Business Intelligence and Technology* strategies and solutions to drive business value. Proven ability to *define* and *drive technology* solutions and services with solid track record of high performing standards and excellence. Has good *scientific* ability and *experience* in *R&D* of Information Control Knowledge based Systems, Planning and Decision-making using business strategy methods in various subject areas using methods of *Situational Control*, *Linguistics* and *Artificial Intelligence*.

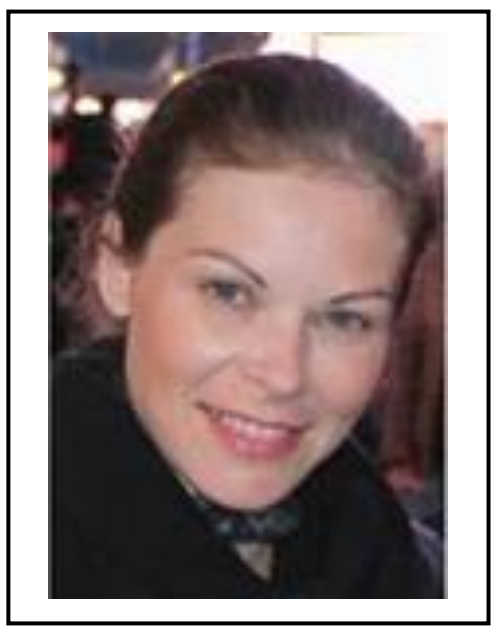

*Maya Abukhana* graduated *Information systems and Management* at the College of Business Management. Have 9 Years of successfully leading and executing Business Projects of Financial and Trading systems to drive business value. Proven ability to *define* and *drive technology* solutions and services with solid track record of high performing standards and excellence. Has a good scientific ability and experience in R&D in the design of knowledge bases, architecture and algorithms for complex financial control systems, as well methods of human interaction using Linguistics and Database technologies.

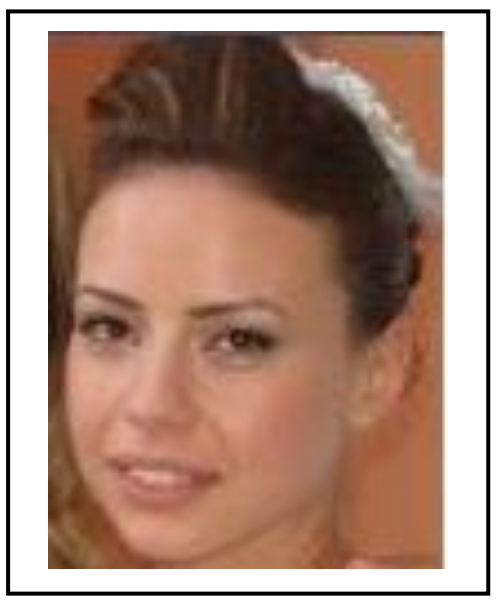